\documentclass[a4paper,10pt]{article}

\usepackage{authblk}

\usepackage[utf8]{inputenc}
\usepackage[T1]{fontenc}

\usepackage{amsmath, amsthm, amssymb}
\usepackage{mathtools}
\usepackage{enumerate}
\usepackage{afterpage}
\usepackage{tabularx}
\usepackage{dsfont}
\usepackage{graphicx}
\usepackage[colorlinks,citecolor=blue,linkcolor=red]{hyperref}

\usepackage{algorithm}
\usepackage{algpseudocode}
\usepackage{caption}
\usepackage[labelformat=simple]{subcaption}

\usepackage{xcolor}
\usepackage{array, multirow}
\usepackage{booktabs}

\newcolumntype{M}[1]{>{\vspace{3pt}\raggedleft\arraybackslash}m{#1}}

\usepackage{siunitx}

\usepackage{cite}

\usepackage{anysize}
\marginsize{2.5cm}{2.5cm}{2cm}{2cm}

\usepackage{wrapfig}

\theoremstyle{plain}

\theoremstyle{remark}

\numberwithin{equation}{section}

\newcommand{\R}{\mathds{R}}
\newcommand{\A}{\mathds{A}}
\newcommand{\B}{\mathds{B}}
\newcommand{\C}{\mathds{C}}

\newcommand{\GG}{\mathds{G}}
\newcommand{\N}{\mathds{N}}

\newcommand{\Sym}[2]{\textrm{Sym}_{#2}(#1)}

\DeclareMathOperator{\Id}{\textrm{Id}}

\newcommand{\fotFourSet}[1]{\texttt{A}(#1)}
\newcommand{\candidateSet}[1]{\texttt{Cand}(#1)}
\newcommand{\QSet}[1]{\texttt{Q}(#1)}
\newcommand{\PositiveSet}[1]{\texttt{N}(#1)}

\newcommand{\specialorthogonalgroup}[1]{\text{SO}(#1)}
\newcommand{\symmetryclass}[1]{\mathcal{S}(#1)}
\newcommand{\symmetryclassspecial}[1]{\mathcal{S}^{#1}}

\newcommand{\completesymmatrix}{
	\multicolumn{3}{c|}{\text{completely}}          & \multicolumn{3}{c}{\text{symmetric}}
}
\usepackage{placeins}
\usepackage{authblk}

\graphicspath{
	{./images/}
}

\title{On the phase space of fourth-order fiber-orientation tensors}

\author[1]{Julian Karl Bauer}
\author[1]{Matti Schneider}
\author[1]{Thomas Böhlke}

\affil[1]{Karlsruhe Institute of Technology (KIT), Institute of Engineering Mechanics, \texttt{\{julian.bauer,matti.schneider,thomas.boehlke\}@kit.edu}}

\date{\today}

\begin{document}

\maketitle

\begin{abstract}
	\noindent {Fiber-orientation tensors describe the relevant features of the fiber-orientation distribution compactly and are thus ubiquitous in injection-molding simulations and subsequent mechanical analyses.}
	In engineering applications to date,
	the second-order fiber-orientation tensor is the basic quantity of
	interest, and the fourth-order fiber-orientation tensor is obtained via a
	closure approximation. Unfortunately, such a description limits
	the predictive capabilities of the modeling process {significantly}, because the wealth of
	possible fourth-order fiber-orientation tensors is not exploited by such
	closures, and the {restriction} to second-order fiber-orientation tensors implies artifacts.
	Closures based on the second-order fiber-orientation tensor face a fundamental
	problem - which fourth-order fiber-orientation tensors can be realized?
	In the literature, only necessary conditions
	for a fiber-orientation tensor to be connected
	to a fiber-orientation distribution
	are found.
	In this article, we show that the typically considered necessary conditions, {positive semidefiniteness and a trace condition},
	are also \emph{sufficient} for being a fourth-order
	fiber-orientation tensor in the physically relevant case of two and three
	spatial dimensions. Moreover, we show that these conditions are not sufficient
	in higher dimensions. The argument is based on convex duality and a celebrated
	theorem of D. Hilbert (1888) on the decomposability of {positive and homogeneous polynomials of degree four}.
	The result has numerous implications for modeling the flow and the resulting
	microstructures of fiber-reinforced composites{, in particular for the effective elastic constants of such materials}.
\end{abstract}

\newpage

\section{Introduction}
\label{sec:intro}

\subsection{State of the art}

Fiber-orientation tensors~\cite{Tucker2022} date back to the far-reaching works~\cite{Kanatani1984, AdvaniTucker1987} and describe the relevant features of the fiber-orientation distribution of discontinuous fiber-reinforced composites.
Within the virtual development and design process of such composites~\cite{Boehlke2019, Goerthofer2019, Tucker2022, Meyer2022}, fiber-orientation tensors appear in material modeling~\cite{Zysset1995, Jack2007, Jack2008, Goldberg2017, Koebler2018, Hessman2021}, microstructure generation~\cite{Feder1980, Pan2008, Altendorf2011, Salnikov2015, Tian2015, Schneider2017, Mehta2022}, mold filling or flow simulations~\cite{Jeffery1922, Folgar1984, Advani1990, Yamamoto1993, Wang2008, Phelps2009} and the experimental computer tomography analysis~\cite{Bay1992, Clarke1995, Geusebroek2003}.
This wide field of application motivates a detailed understanding of the mathematical properties of fiber-orientation tensors.
{Actually,} the motivation and field of application is {much more general}, as fabric tensors and diffusion tensors share structural properties with fiber-orientation tensors.
Fabric tensors~\cite{Cowin1985,Zysset1995} share all characteristics with fiber-orientation tensors, except for a normalization constraint and are used in the field of porous materials.
Diffusion tensors~\cite{Le2001, Mori2002, Basser2002} also differ from fiber-orientation tensors only
by a missing trace constraint and
are used in medicine to describe the orientation of
body tissues
based on the diffusion motion of water molecules.
This procedure is called diffusion-weighted magnetic resonance imaging (DW-MRI)~\cite{Taylor1985, Le1986}
and is, e.g., used on brain tissue to prevent strokes~\cite{Schaefer2000}.
In particular, insights on the mathematical properties of fiber-orientation tensors might be transferred to
diffusion tensors or fabric tensors.

The phase space of second-order fiber-orientation tensors is known
\cite{Nomura1970, Cintra1995, Chung2001, Linn2005, Mueller2016}, {see the recent review~\cite{Bauer2022Variety}}.
A spectral decomposition is typically used
to separate the structural features of a second-order fiber-orientation tensor, described by
two independent eigenvalues with limited admissible parameter ranges, from
the spatial alignment of this structural information in terms of a rotation or eigensystem.
The limited structural variability of second-order fiber-orientation tensors
is a critical ingredient for applications in process simulations
\cite{Jack2008, Goldberg2017, Koebler2018, Goerthofer2019}.

In contrast to the second-order case,
the phase space of fourth-order fiber-orientation tensors
is less well understood, a circumstance which motivated the work at hand.
Algebraic properties of symmetry and normalization are agreed upon and
discussed in {the literature}~\cite{Moakher2009, Rahmoun2009, Weldeselassie2012, Ghosh2012, Moreno2016}.
Bauer and Böhlke~\cite{Bauer2022Variety}
develop{ed} an eigensystem-based parameterization
of fourth-order fiber-orientation tensors
combining the framework of
irreducible tensors
\cite{Spencer1970, Forte1996, Jerphagnon1978, Adams1992, Cowin1989, Rychlewski2000, Bauer2023Diss}
with the work of Kanatani~\cite{Kanatani1984}.
This parameterization ensures normalization {as well as} symmetry conditions {automatically} and
separates second- and fourth-order data.
{Moreover, additional material-symmetry constraints} may be
taken into account in a natural way.
Bauer and Böhlke~\cite{Bauer2022Variety}
assume that positive semidefiniteness of
the tensor
is a sufficient condition to derive
admissible parameter ranges, specifying
the variety of fourth-order fiber-orientation tensors.
This variety is presented for special cases motivated by material symmetry.
The case of planar fourth-order fiber-orientation tensors and derived quantities is studied in
successive papers~\cite{Bauer2022_fiber, Bauer2022_onthe}.
However, the necessary condition of positive semidefiniteness
of the completely symmetric tensor is assumed to be sufficient.
%
In section \ref{sec:fot_suff} of this work, sufficiency of this condition for the
cases inspected by Bauer and Böhlke~\cite{Bauer2022Variety} is proven.

A scientific topic which is intimately connected to the question on the
phase space of fourth-order fiber-orientation tensors,
is fiber-orientation closure approximations.
Such closure approximations
\cite{FolgarTucker1984, Advani1990, Han1999, Cintra1995, Chung2002,
Montgomery2011, Montgomery2011thefast, Karl2021_2, Tucker2022_planar}
are tensor-valued functions
which postulate a functional relationship
between a given second-order fiber-orientation tensor
and an unknown fourth-order fiber-orientation tensor~\cite{Mehta2022}.
Identifying the phase space of fourth-order fiber-orientation tensors
is essential to solve a fundamental problem of closure approximations
- which fourth-
order fiber-orientation tensors can be realized?

\subsection{Contributions}
This work is divided into a {basic} and {an applied} part.
{In the first part, we provide a proof for a simple characterization for the phase space of realizable fourth-order fiber-orientation tensors in the physically relevant dimensions two and three. In case of second-order fiber-orientation tensors, such a characterization is straightforward to obtain. Any symmetric second-order tensor whose eigenvalues are non-negative and sum to unity represents such a fiber-orientation tensor. In case of fourth-order tensors, an educated guess gives that completely symmetric fourth-order tensors whose eigenvalues are non-negative and sum to unity should correspond to realizable fourth-order fiber-orientation tensors.\\
Our arguments are based on two key ingredients. First, we use convex duality theory, in the following form. Two closed convex cones coincide precisely if their dual cones coincide. For the case at hand, the dual cones are a bit simpler to work with. The second insight is based on the natural} identification of completely symmetric tensors {of order $k$} with homogeneous polynomials {of degree $k$. Then, a celebrated theorem due to Hilbert~\cite{Hilbert1888} on} the decomposability of a {positive} homogeneous polynomial of {fourth degree} into a sum of squares {of homogeneous quadratic polynomials implies the claim. Interestingly, this theorem holds only in spatial dimensions two and three. For higher dimensions, this decomposability does not hold. In particular, the educated guess for characterizing the phase-space of fourth-order fiber-orientation tensors is actually \emph{false} for spatial dimensions four and above. This fact shows that there are no "elementary", in particular dimension-independent, arguments for showing this characterization of the phase space to be valid.\\
As a by-product of our analysis, we show the} equivalence of representations of fiber-orientation tensors in terms of integration, sums of monomials (rank-one tensors) {and} finite sums of monomials with non-negative weights summing to one, {a consequence of} Carathéodory's theorem~\cite{Caratheodory1911}.\\
In the {applied} part, {we investigate the geometry of the} phase space of fourth-order fiber-orientation tensors {via studying}
optimization {problems on the latter. For this purpose, we consider suitable semidefinite programs, i.e., optimization problems posed on positive semidefinite matrices. Indeed, due to the characterization shown in the first part, we may identify a realizable fourth-oder fiber-orientation tensor with a positive semidefinite matrix in a suitable matrix representation (we use the Kelvin-Mandel notation).}\\
Extreme states {of} the phase space of fourth-order fiber-orientation tensors
are studied based on {the parameterizations introduced in previous work}~\cite{Bauer2022Variety},
which separate second- and fourth-order data and enable {to incorporate} constraints on material symmetry {easily}.
For fixed second-order information, we identify fourth-order data maximizing the projection onto
a specified direction in two and three dimensions.
Restricting the parameter space to orthotropic fourth-order information
prohibits extreme projections into certain directions.
{These findings} demonstrate {the inherent limitations} of closure approximations,
as closed fourth-order information is automatically orthotropic. {Due to the close connection to effective elastic properties, corresponding limitations of the effective stiffnesses become clear, as well.}
%


This paper is organized as follows.
After recapitulating basic properties of fiber-orientation tensors
and {their} representation {by} finite sums, see section \ref{sec:fot_def},
we specify a set of fiber-orientation tensor candidates in section \ref{sec:fot_nec}.
In section \ref{sec:fot_suff}, we establish sufficiency of the conditions these candidates fulfill
for tensor order four and dimensions two and three,
leading to applications in section \ref{sec:consequences}.
In section \ref{sec:consequences_convex_optimization},
{optimization} on fiber-orientation tensors is identified as semidefinite programming.
Consequences of material symmetry constraints, introduced in section \ref{sec:material_symmetry_constraints},
on the problem of extreme fourth-order information,
defined in section \ref{sec:extreme_fourth_order_information},
are presented in sections \ref{sec:results_3d} and \ref{sec:results_2d} for dimensions three and two.

\newpage

\section{Realizability of fourth-order fiber-orientation tensors}
\label{sec:fot}

\subsection{Fiber-orientation tensors}
\label{sec:fot_def}

A non-polar
fiber-orientation distribution is given by a probability measure\footnote{On
	the $\sigma$-algebra of Borel sets.} $\mu$ on the unit sphere
\begin{equation}
	S^{d-1} = \left\{ p \in \R^d \, \middle| \, \|p\|=1 \right\}
\end{equation}
in $d$ spatial dimensions which is invariant w.r.t. the involution $p\mapsto
	-p$ on the unit sphere. This definition includes continuous fiber-orientation
distributions, described by non-negative and suitably normalized functions
$\varphi$ via the measure 
\begin{equation}
	{\mu = \varphi \, dA}
\end{equation} 
in terms of the usual surface measure
$dA$ on the unit sphere. Moreover, discrete fiber-orientation states, described
in terms of $N$ vectors $p_i \in S^{d-1}$ are included via the formulation
\begin{equation}\label{eq:fot_def_discrete}
	\mu = \frac{1}{2N} \sum_{i=1}^N \left( \delta_{p_i} + \delta_{-p_i}
	\right),
\end{equation}
where $\delta_p$ denotes the Dirac measure concentrated on the point $p$.\\
As working with such measures may be cumbersome,
Advani-Tucker~\cite{AdvaniTucker1987} (compare also
Kanatani~\cite{Kanatani1984}) introduced the so-called fiber-orientation
tensors $A^{(k)}$ of order $k$, {where $k$ denotes} a non-negative integer, via the definition
\begin{equation}\label{eq:fot_def}
	A^{(k)} = \int_{S^{d-1}} p^{\otimes k} \, d\mu(p)
\end{equation}
in terms of the $k$-fold outer tensor product $p^{\otimes k}$ of the unit
vector $p$, integrated over the unit sphere. The fiber-orientation tensors have
the following properties.
\begin{enumerate}
	\item The fiber-orientation tensors are completely symmetric, i.e., the
	      identity
	      \begin{equation}\label{eq:fot_def_complete_symmetry_naive}
		      A^{(k)}_{i_1 \, i_2 \, \ldots \, i_k} = A^{(k)}_{i_{\sigma(1)}
		      \, i_{\sigma(2)} \, \ldots \, i_{\sigma(k)}}, \quad i_j \in \{1,\ldots,d\},
		      \quad j\in \{1,\ldots,k\}
	      \end{equation}
	      holds for the components of the tensor $A^{(k)}$ and any permutation
	      $\sigma$ of the index set $\{1,\ldots,k\}$. Denoting by $\Sym{k}{d}$ the vector
	      space of completely symmetric $k$-th order tensors in $d$ dimensions, the
	      condition \eqref{eq:fot_def_complete_symmetry_naive} is equivalent to the
	      inclusion $A^{(k)} \in {\Sym{d}{k}}$.
	\item For odd order $k$, the fiber-orientation tensors vanish.
	\item The fiber-orientation tensor $A^{(k)}$ is positive semidefinite
	      in the sense that
	      \begin{equation}\label{eq:fot_def_nonnegative_onefold}
		      A^{(k)} \cdot^k q^{\otimes k} \geq 0
	      \end{equation}
	      holds for all vectors $q \in \R^d$ and the $k$-fold index contraction
	      $\cdot^k$ of tensors.
	\item Double contraction with the $d \times d$-identity $\Id$ recovers
	      the fiber-orientation tensor of two orders lower
	      \begin{equation}\label{eq:contract_to_lower_order}
		      A^{(k+2)}:\Id = A^{(k)}, \quad k = 0,1,2,\ldots
	      \end{equation}
	\item The zeroth-order fiber-orientation tensor is unity, i.e., the
	      equation
	      \begin{equation}\label{eq:fot_zeroth_one}
		      A^{(0)} = 1
	      \end{equation}
	      holds.
	\item Any fixed fiber-orientation tensor $A^{(k)}$ of even order may be
	      written as a finite sum
	      \begin{equation}\label{sec:fot_def_monomial_decomposition}
		      A^{(k)} = \sum_{i=1}^r w_i \, q_i^{\otimes k}
	      \end{equation}
	      with appropriate non-negative weights $w_i$ which sum to unity,
	      directions $q_i \in S^{d-1}$ and a non-negative integer $r$ not exceeding
	      ${{k+d-1}\choose{d-1}}$.
\end{enumerate}
These {properties} 1.-5. are elementary to {verify} and
well-known~\cite{AdvaniTucker1987}. We provide a derivation of property 6. in
the Appendix \ref{sec:atomic_decomposition_Ak}. Property 6. permits us to
restrict to finite sums \eqref{sec:fot_def_monomial_decomposition} when
considering fiber-orientation tensors of fixed order. In particular, it is not
necessary to resort to the general integration-based definition
\eqref{eq:fot_def}.

\subsection{Fiber-orientation tensor candidates}
\label{sec:fot_nec}

The work at hand is specifically concerned with fourth-order fiber-orientation
tensors $A^{(4)} \equiv \A$. {More precisely, we are interested in the set of realizable fourth-order fiber-orientation tensors
\begin{equation}
	\fotFourSet{d} = \left\{ \A \in \Sym{d}{4} \, \middle| \, \A = \int_{S^{d-1}} p^{\otimes 4} \, d\mu(p) \quad \text{for a fiber-orientation distribution} \quad \mu \right\}.
\end{equation} 
By Property 6. of the previous section, the set $\fotFourSet{d}$ may be rewritten in the form
\begin{equation}\label{sec:fot_nec_fot4_defn}
	\fotFourSet{d} = \left\{ \A = \sum_{i=1}^r \mu_i \, p_i^{\otimes 4} \, \middle| \, r>0, \quad {\mu_i > 0} \quad \text{with} \quad \sum_{i=1}^r \mu_i = 1, \quad p_i \in S^{d-1}\right\},
\end{equation}
which is more convenient from the mathematical point of view.}\\
The fundamental question we wish to answer in this work is the following.
Suppose a fourth-order tensor $\A$ is given - under which (easy to verify)
conditions can it be represented in the form \eqref{sec:fot_nec_fot4_defn}?
Here, the non-negativity of the weights is crucial. Indeed, every completely
symmetric $k$-th order tensor can be written as a linear combination of
monomials~\cite[Lemma 4.2]{Comon2008}{, i.e., with both positive \emph{and} negative weights.\\
Elements $\A \in \fotFourSet{d}$ are completely symmetric \eqref{eq:fot_def_complete_symmetry_naive},
positive semidefinite {on quartics} \eqref{eq:fot_def_nonnegative_onefold}
\begin{equation}\label{eq:fot_def_nonnegative_onefold_explicit}
	\A:: q^{\otimes 4} \geq 0 \quad \text{for all} \quad q \in \R^d,
\end{equation}
and satisfy the trace-conditions \eqref{eq:contract_to_lower_order} {as well as} \eqref{eq:fot_zeroth_one}. Here, the operator $::$ stands for a fourth-fold contraction, and gives rise to an inner product on the space $\Sym{d}{4}$ of completely symmetric fourth-order tensors.\\
However, it is readily seen that fiber-orientation tensors $\A \in \fotFourSet{d}$ satisfy a non-negativity condition which is \emph{stronger} than the condition \eqref{eq:fot_def_nonnegative_onefold_explicit}. More precisely, for any tensor $\A \in \fotFourSet{d}$, the inequality
\begin{equation}\label{sec:fot_nec_nonnegative_twofold}
	S:\A:S \geq 0 \quad \text{holds for all} \quad S \in \Sym{d}{2},
\end{equation}
i.e., symmetric {second-order} tensors $S$ {in dimension $d$}. Indeed, in view of the representation
\eqref{sec:fot_nec_fot4_defn}, we observe
\begin{equation}
	S:\A:S = \sum_{i=1}^r \mu_i \, S:p_i^{\otimes 4}:S = \sum_{i=1}^r \mu_i
	\, \left(S:p_i^{\otimes 2}\right)^2 \geq 0
\end{equation}
for any $S \in \Sym{d}{2}$ due to the non-negativity of the weights $\mu_i$. Please notice that the condition
\eqref{sec:fot_nec_nonnegative_twofold} is stronger than the condition \eqref{eq:fot_def_nonnegative_onefold_explicit} as the special symmetric tensor $S = q \otimes q$ may be chosen in the
condition \eqref{sec:fot_nec_nonnegative_twofold}. Moreover, the condition \eqref{sec:fot_nec_nonnegative_twofold} is equivalent to the positive semidefiniteness of the tensor $\A$, considered as a linear operator 
\begin{equation}
	\Sym{d}{2} \rightarrow \Sym{d}{2}, \quad S \mapsto \A:S,
\end{equation}
on the vector space $\Sym{d}{2}$, endowed with the inner product $(S,T)\mapsto S:T$.}\\
In the literature, the following set is typically
considered~\cite{Bauer2022Variety}
\begin{equation}\label{sec:fot_nec_candidates}
	\candidateSet{d} = \left\{ \A \in \Sym{d}{4} \, \middle| \, \A \quad
	\text{{is positive semidefinite} \eqref{sec:fot_nec_nonnegative_twofold}}, \quad \Id:\A:\Id =
	1\right\}
\end{equation}
to consist of reasonable candidates for fiber-orientation tensors.\\
The set \eqref{sec:fot_nec_candidates} is much easier to
work with {in practice} than the original set $\fotFourSet{d}$. Indeed, suppose a candidate
tensor $\A$ is given. Then, it is elementary to {examine} its complete symmetry.
Moreover, the non-negativity condition \eqref{sec:fot_nec_nonnegative_twofold}
and the normalization requirement $\A:: \Id \otimes \Id = 1$ are {assessed} via an eigendecomposition of the fourth-order tensor $\A${, e.g., conveniently computed in an orthogonal basis like the Kelvin-Mandel
representation~\cite{Thomson1856, Mandel1965}}. Indeed, both
conditions are satisfied precisely if the eigenvalues are non-negative and sum
to unity.\\
In contrast, the defining condition {of realizable fiber-orientation tensors} \eqref{sec:fot_nec_fot4_defn} is
much harder to check, as it involves a canonical decomposition problem with
non-negativity constraints, which is known to be
non-trivial~\cite{Comon2008}.\\
{Properties 1., 4.
and 5. of fiber-orientation tensors, listed in the previous section
\ref{sec:fot_def}, imply all defining conditions of the set
$\candidateSet{d}$. Moreover, we just established the positive semidefiniteness condition \eqref{sec:fot_nec_nonnegative_twofold}. Thus, the inclusion of sets
}
\begin{equation}\label{sec:fot_nec_set_inclusion}
	\fotFourSet{d} \subseteq \candidateSet{d}
\end{equation}
holds. The impending question remains whether the set inclusion \eqref{sec:fot_nec_set_inclusion} is strict or not{, i.e., whether every element $\A \in \candidateSet{d}$ is also realizable, i.e., $\A \in \fotFourSet{d}$ holds.}

\subsection{On sufficiency of conditions}
\label{sec:fot_suff}

The purpose of this section is to check when the inverse direction of the
inclusion \eqref{sec:fot_nec_set_inclusion}
\begin{equation}\label{sec:fot_suff_set_inclusion}
	\candidateSet{d} \subseteq \fotFourSet{d}
\end{equation}
holds. It will turn out that this is true for spatial dimensions $d \leq 3$ and
false for dimension $d \geq 4$.\\
{Central to our arguments is the auxiliary} set
\begin{equation}\label{sec:fot_suff_QSet}
	\QSet{d} = \left\{ \A \in \Sym{d}{4} \,\middle|\, \A = \sum_{i=1}^r
	\mu_i \, p_i^{\otimes 4}, \quad \mu_i \in \R_{\geq 0}, \quad p_i \in S^{d-1},
	\quad r\in \N_0 \right\}
\end{equation}
of sums of fourth-order tensor-powers of vectors {with non-negative weights} and the set of tensors
\begin{equation}\label{sec:fot_suff_PositiveSet}
	\PositiveSet{d} = \left\{ \A \in \Sym{d}{4} \,\middle|\, S:\A:S \geq 0
	\quad \text{for all} \quad S \in \Sym{d}{2} \right\}
\end{equation}
which {are completely symmetric and positive semidefinite} \eqref{sec:fot_nec_nonnegative_twofold}. {Notice that, in contrast to the definition \eqref{sec:fot_nec_fot4_defn} of the set $\fotFourSet{d}$, we use non-negative weights in the definition \eqref{sec:fot_suff_QSet} instead of positive weights to include the zero tensor in the set $\QSet{d}$. This inclusion is necessary for the set $\QSet{d}$ to be \emph{closed}.}\\
The intersection conditions
\begin{equation}\label{sec:fot_suff_intersection_condition1}
	\fotFourSet{d} = \QSet{d} \cap \left\{ \A \in \Sym{d}{4} \,\middle|\,
	\Id:\A:\Id = 1\right\}
\end{equation}
and
\begin{equation}\label{sec:fot_suff_intersection_condition2}
	\candidateSet{d} = \PositiveSet{d} \cap \left\{ \A \in \Sym{d}{4}
	\,\middle|\, \Id:\A:\Id = 1\right\}
\end{equation}
hold. Thus, we have the implication
\begin{equation}
	\text{if} \quad \QSet{d} = \PositiveSet{d} \quad \text{holds, then}
	\quad \fotFourSet{d} = \candidateSet{d} \quad \text{follows.}
\end{equation}
In particular, the fiber-orientation realization problem
\eqref{sec:fot_suff_set_inclusion} may be studied in terms of the sets
\eqref{sec:fot_suff_QSet} and \eqref{sec:fot_suff_PositiveSet}. Both sets are
closed convex cones{, i.e., they are closed, convex and invariant under the action $\A \mapsto \lambda \A$ for $\lambda \geq 0$. For the set $\PositiveSet{d}$, the definition \eqref{sec:fot_suff_PositiveSet} immediately implies that it is a closed convex cone. Indeed, inequality constraints involving continuous functions lead to closed sets in finite dimensions, and the representation $S:\A:S = \A:: {\left(S \otimes S\right)}$ recasts the set \eqref{sec:fot_suff_PositiveSet} as an intersection of half spaces through zero. As the intersection of (a family of) convex cones is a convex cone, we see that the set \eqref{sec:fot_suff_PositiveSet} is a convex cone}.\\
 For the set $\QSet{d}$, the definition \eqref{sec:fot_suff_QSet} {implies} that
the set is a convex cone. The closedness is a consequence of Carathéodory's
theorem~\cite{Caratheodory1911}, as detailed in the Appendix \ref{sec:atomic_decomposition_Ak}.
Indeed, as a consequence of this theorem, we may
assume that no more than $r_{\max} = {{k+d-1}\choose{d-1}}$ terms enter in the
definition \eqref{sec:fot_suff_QSet}. Suppose a sequence $\left(\A_k\right)$ of
elements in $\QSet{d}$ converges to some $\A \in \Sym{d}{4}$. Let us write
\begin{equation}
	\A_k = \sum_{i=1}^{r_{\texttt{max}}} \mu_{i,k} \, p_{i,k}^{\otimes 4},
	\quad \mu_{i,k}\geq 0, \quad p_{i,k} \in S^{d-1}.
\end{equation}
We observe
\begin{equation}
	\sum_{i=1}^{r_{\texttt{max}}} \mu_{i,k} = \Id:\A_k:\Id \rightarrow
	\Id:\A:\Id \quad \text{as} \quad k \rightarrow \infty.
\end{equation}
Thus, there is a uniform bound $C$ on the coefficients $\mu_{i,k}$. As the sets
$[0,C]$ and $S^{d-1}$ are compact, we may extract a subsequence (not relabeled), s.t.
\begin{equation}
	\mu_{i,k} \rightarrow \mu_i \quad \text{and} \quad p_{i,k} \rightarrow
	p_i \quad \text{as} \quad k\rightarrow \infty
\end{equation}
for suitable $\mu_i \in [0,C]$ and $p_i \in S^{d-1}$. Due to the uniqueness of
limits, we obtain the representation
\begin{equation}
	\A = \sum_{i=1}^{r_{\texttt{max}}} \mu_i\,p^{\otimes 4}, \quad
	\text{i.e.}, \quad \A \in \QSet{d}.
\end{equation}
Hence, the set $\QSet{d}$ is closed.\\
For any set $C \subseteq \R^m$ with inner product $\langle \cdot,\cdot
	\rangle$, the dual cone $C^*\subseteq \R^m$ is defined by the relation
\begin{equation}\label{sec:fot_suff_dual_cone_def}
	C^* = \left\{ z \in \R^m\,\middle|\, \langle z, x \rangle \geq 0 \quad
	\text{for all} \quad x \in C \right\}.
\end{equation}
It is immediate to see that {the set} $C^*$ does indeed define a closed convex cone. If,
moreover, the original set $C$ was a closed convex cone to start with, the dual
cone of the dual cone coincides with the original set
\begin{equation}
	C^{**} = C.
\end{equation}
This assertion is a consequence of convex duality theory~\cite{Rockafellar}.\\
As we identified the sets $\QSet{d}$ and $\PositiveSet{d}$ as closed convex
cones, the equivalence
\begin{equation}
	\QSet{d} = \PositiveSet{d} \quad \text{precisely if} \quad \QSet{d}^* =
	\PositiveSet{d}^*
\end{equation}
follows directly, where we fix the inner product
\begin{equation}
	\langle \A,\B\rangle = \A::\B, \quad \A,\B \in \Sym{d}{4}.
\end{equation}
For this assertion to be useful, it is necessary to identify the dual cones
explicitly. We have
\begin{equation}\label{sec:fot_suff_QSet_dual}
	\QSet{d}^* = \left\{ \B\in\Sym{d}{4} \,\middle|\, \B::q^{\otimes k}
	\geq 0 \quad \text{for all} \quad q \in \R^d \right\}.
\end{equation}
Indeed, the inclusion $\subseteq$ follows from the definition of the dual cone
\eqref{sec:fot_suff_dual_cone_def}
\begin{equation}
	\QSet{d}^* = \left\{ \B \in \Sym{d}{4} \, \middle| \, \A::\B \geq 0
	\quad \text{for all} \quad \A \in \QSet{d} \right\}
\end{equation}
by selecting $\A = p^{\otimes 4}$ as an element of $\QSet{d}$. The inclusion
$\supseteq$ follows by considering
\begin{equation}
	\A::\B = \sum_{i=1}^r \mu_i \, \underbrace{B::p_i^{\otimes 4}}_{\geq 0}
	\geq 0 \quad \text{for} \quad \A \in \QSet{d}.
\end{equation}
With a completely analogous argument, we obtain
\begin{equation}\label{sec:fot_suff_PositiveSet_dual}
	\PositiveSet{d}^* = \left\{ \B \in \Sym{d}{4} \, \middle| \, \B =
	\sum_{i=1}^r \texttt{sym}(S_i \otimes S_i), \quad S_i \in \Sym{d}{2} \right\},
\end{equation}
where $\texttt{sym}$ refers to the complete symmetrization of all indices.
To study when the sets $\QSet{d}^*$ and $\PositiveSet{d}^*$ coincide, we use the
identification of completely symmetric tensors with homogeneous polynomials.
More precisely, for any {tensor} $\B \in \Sym{d}{4}$, the function
\begin{equation}\label{sec:fot_suff_hom_poly}
	P_{\B}: \R^d \ni x \mapsto \B::x^{\otimes 4} \in \R
\end{equation}
defines a homogeneous polynomial of order four. Conversely, any fourth-order
homogeneous polynomial may be represented in the form
\eqref{sec:fot_suff_hom_poly}. With this interpretation, {a tensor} $\B \in \Sym{d}{4}$ is
an element of $\QSet{d}^*$ (see equation \eqref{sec:fot_suff_QSet_dual}) precisely if
the polynomial $P_{\B}$ is non-negative
\begin{equation}\label{sec:fot_suff_positive_polynomial}
	P_{\B}(x) \geq 0 \quad \text{for all} \quad x \in \R^d.
\end{equation}
Similarly, a tensor $\B \in \Sym{d}{4}$ is an element of the set
$\PositiveSet{d}^*$
\eqref{sec:fot_suff_PositiveSet_dual} precisely if the associated polynomial is
a sum of squares
\begin{equation}\label{sec:fot_suff_sos_polynomial}
	P_{\B}(x) = \sum_{i=1}^r Q_i(x)^2
\end{equation}
with homogeneous quadratic polynomials $Q_i$. The decomposability of a positive
homogeneous polynomial of even order \eqref{sec:fot_suff_positive_polynomial}
into a sum of squares \eqref{sec:fot_suff_sos_polynomial} is a classical
problem of real algebraic geometry, first addressed by
Hilbert~\cite{Hilbert1888}. In our terminology, Hilbert's results imply that
the identity
\begin{equation}
	\QSet{d}^* = \PositiveSet{d}^*
\end{equation}
holds for dimensions $d\leq 3$ and is false otherwise. Explicit
counterexamples~\cite{Counterexample1,Counterexample2} and a quantitative
version~\cite{Blekherman2006} are available.\\
Thus, we obtain the assertion
\begin{equation}
	\fotFourSet{d} = \candidateSet{d}
\end{equation}
in the physically relevant cases $d=2$ and $d=3$. For higher dimensions, these
assertions are false as a result of Hilbert's results~\cite{Hilbert1888} and
the fact that the sets $\candidateSet{d}$ and $\fotFourSet{d}$ arise via a
codimension-one condition, see equations
\eqref{sec:fot_suff_intersection_condition1} and
\eqref{sec:fot_suff_intersection_condition2}.

\section{Consequences and variations}
\label{sec:consequences}

\subsection{Optimal microstructure: Programming on fiber-orientation tensors}
\label{sec:consequences_convex_optimization}

Identifying microstructures which lead to extreme effective physical properties
of microstructured composite materials is a central goal in the field of material design.
Following Section \ref{sec:fot_suff},
and focusing on
the practically relevant case of fourth-order fiber-orientation tensors in two or three dimensions,
i.e., $d=2$ or $d=3$  and $k=4$,
the condition
$
	\A \in \candidateSet{d}
$
with \eqref{sec:fot_nec_candidates}
\begin{equation}
	\candidateSet{d}
	=
	\left\{ \A \in \Sym{d}{4} \, \middle| \, \quad
	{\A} \in \PositiveSet{d},
	\quad \Id:\A:\Id =
	1\right\}
\end{equation}
is sufficient for the tensor $\A$ to be realizable,
i.e., there exists at least one microstructure composed of fibers
which is described by {the tensor} $\A$ on average.
In consequence, one strategy to identify extreme microstructures is to optimize w.r.t. fiber-orientation tensors.
Optimization on the phase space of fourth-order fiber-orientation tensors
may be accomplished via semidefinite programming,
as the set
\begin{equation}
	\PositiveSet{d} = \left\{ \A \in \Sym{d}{4} \,\middle|\, S:\A:S \geq 0
	\quad \text{for all} \quad S \in \Sym{d}{2} \right\},
\end{equation}
defined in equation \eqref{sec:fot_suff_PositiveSet},
is a closed convex cone and requires {the tensor} $\A$ to be positive semidefinite.
The normalization or trace condition $\Id:\A:\Id {=1}$
as well as the complete symmetry condition $\A \in \Sym{d}{4}$
represent linear constraints of the semidefinite program.
Linear semidefinite programming on the phase space of fourth-order fiber-orientation tensors
solves the problem
\begin{equation}
	\begin{array}{ll}
		\text{maximize}  		& \C :: \A         \\
		\text{subject to}		& \A \in \candidateSet{d} 	\\
								& \GG_k :: \A = g_k, \quad k = 1, ..., m,
	\end{array}
	\label{eq:semidefinite_program_tensor}
\end{equation}
with the fourth-order
tensor
$\C$ entering the objective function and
tuples $(\GG_k, g_k)$
of fourth-order
tensors
and scalars
specifying $m$ additional linear constraints.
In order to solve the problem \eqref{eq:semidefinite_program_tensor},
a transition from tensorial notation to the Kelvin-Mandel notation is advised.
In Kelvin-Mandel notation, fourth-order tensors with the left and right minor symmetry are represented
in an orthonormal basis of symmetric second-order tensors.
In contrast to the commonly used Voigt notation~\cite{Voigt1910}, the Kelvin-Mandel basis tensors are not only orthogonal but also normalized.
As a result, in dimensions two and three, the eigenvalues of the matrix with respect to the Kelvin-Mandel basis tensors
are identical to the eigenvalues of a fourth-order tensor with the aforementioned symmetries.
The change from tensorial notation to Kelvin-Mandel notation is made possible by the symmetries of the problem and
allows to reduce the dimension of the problem by introducing tensor bases.
In the case $d=3$, $k=4$ and the symmetries at hand,
it is possible to represent the tensor
\begin{equation}
	\A = \sum_{\xi=1}^6 \sum_{\zeta=1}^6 A_{\xi \zeta} \, \mathbf{B}_{\xi} \otimes \mathbf{B}_{\zeta},
\end{equation}
in terms of a six times six coefficient matrix $\A_{\xi \zeta}$
and the dyads $\mathbf{B}_{\xi} \otimes \mathbf{B}_{\zeta}$
with the indices $\xi, \zeta \in \{1,\,...,\, 6\}$.
Details of the Kelvin-Mandel notation are given in
Bauer and Böhlke~\cite{Bauer2022Variety}.
The complete index symmetry of {the tensor} $\A$ stated in equation \eqref{eq:fot_def_complete_symmetry_naive}
is easily imposed by linear constraints,
since the representation of a completely symmetric tensor $\N$ in Kelvin-Mandel notation reads
\definecolor{newgreen}{HTML}{2CA918}
\begin{align}
	\N
	 & =
	\left[
		\begin{tabular}{lll|lll}
			$N^{(4)}_{11}$                              & $N^{(4)}_{\color{blue}12}$ & $N^{(4)}_{ \color{olive}13}$   & $\sqrt{2} N^{(4)}_{\color{purple}14}$ & $\sqrt{2}N^{(4)}_{15}$               & $\sqrt{2}N^{(4)}_{16}$             \\
			                                            & $N^{(4)}_{22}$             & $N^{(4)}_{\color{newgreen}23}$ & $\sqrt{2}N^{(4)}_{24}$                & $\sqrt{2}N^{(4)}_{\color{orange}25}$ & $\sqrt{2}N^{(4)}_{26}$             \\
			                                            &                            & $N^{(4)}_{33}$                 & $\sqrt{2}N^{(4)}_{34}$                & $\sqrt{2}N^{(4)}_{35}$               & $\sqrt{2}N^{(4)}_{\color{teal}36}$ \\\hline
			                                            &                            &                                & $2N^{(4)}_{\color{newgreen}23}$       & $2N^{(4)}_{\color{teal}36}$          & $2N^{(4)}_{\color{orange}25}$      \\
			\multicolumn{3}{c|}{\text{symmetric}} &                            & $2N^{(4)}_{\color{olive}13}$   & $2N^{(4)}_{\color{purple}14}$                                                                                     \\
			                                            &                            &                                &                                       &                                      & $2N^{(4)}_{\color{blue}12}$        \\
		\end{tabular}
		\right]
	\mathbf{B}_{\xi} \otimes \mathbf{B}_{\zeta}
	\label{eq:complete_sym},
\end{align}
with redundant coefficients color-coded following the reference~\cite{Bauer2022Variety}
and ``symmetric'' indicating matrix symmetry of the tensor coefficients in equation \eqref{eq:complete_sym}.

An active rotation of a fiber-orientation tensor changes the
alignment in space of the averaged directional information on the fibers' orientation.
However, the averaged information itself, {apart from} its alignment in space,
is unaffected.
If we are interested in studying the averaged information itself, independent of its alignment in space,
we can use an eigensystem-based parameterization of fourth-order fiber-orientation tensors~\cite{Bauer2022Variety}.
The $15$-dimensional space $\Sym{3}{4}$,
reduced by the normalization constraint \eqref{eq:fot_zeroth_one},
leaves $14$ dimensions, i.e., three defining the spatial alignment of the averaged information and eleven remaining dimensions
which represent structural information.
In addition to distinguishing alignment and structural information,
the parameterization of Bauer and Böhlke~\cite{Bauer2022Variety}
separates second- and fourth-order information following the seminal work of Kanatani~\cite{Kanatani1984}.
Therefore, searching for fourth-order data which maximizes the objective function for given second-order information is possible.
Further, material symmetry properties can be easily incorporated
by using the additional constraints in equation \eqref{eq:semidefinite_program_tensor}.
Following Bauer and Böhlke~\cite[Equation (59)]{Bauer2022Variety},
a general, i.e., triclinic, fourth-order fiber-orientation tensor
may be parameterized by
\begin{equation}
	\A
	\left( \mathbf{A},\, d_1,\, ..., \, d_9 \right)
	=
	\A^{\text{iso}}
	+
	\frac{6}{7} \;
	\text{sym}
	\left(
	\text{dev}\left(\mathbf{A}\right)
	\otimes
	\Id
	\right)
	+
	\mathbb{F}^{\text{tricl}}\left( d_1,\, ..., \, d_9\right)
	\label{eq:N4_simplified_triclinic}
\end{equation}
based on an eigensystem-based parameterization of the corresponding second-order fiber-orientation tensor
\begin{equation}
	\mathbf{A}
	=
	\sum_{i=1}^{3}
	\sum_{j=1}^{3}
	\hat{A}_{ij}
	\,
	\mathbf{e}_i \otimes \mathbf{e}_j
	=
	\sum_{i=1}^{3}
	\lambda_i
	\,
	\mathbf{v}_i \otimes \mathbf{v}_i,
\end{equation}
where we apply the common ordering convention
\begin{align}
	\lambda_1 \ge \lambda_2 \ge \lambda_3.
	\label{eq:convention_eigenvalues_N2}
\end{align}
In addition, equation \eqref{eq:N4_simplified_triclinic} contains
the isotropic fourth-order fiber-orientation tensor
\begin{align}
	\A^{\text{iso}}
	=
	\frac{7}{35}
	\text{sym}
	\left(
	\Id \otimes \Id
	\right)
	\label{eq:N4_iso},
\end{align}
a triclinic fourth-order deviatoric structure tensor
represented in Kelvin-Mandel notation by
\begin{align}
	&
	\mathbb{F}^{\text{tricl}}\left( d_1, ..., \, d_9\right)
	=
	\label{eq:triclinic_structure_tensor}\\
	&\quad
	{
	\footnotesize
	\left[
		\begin{array}{lll|lll}
			-(d_1+d_2) & d_1        & d_2        & -\sqrt{2}(d_4+d_5) & \sqrt{2}d_6        & \sqrt{2}d_8         \\
			           & -(d_1+d_3) & d_3        & \sqrt{2}d_4        & -\sqrt{2}(d_6+d_7) & \sqrt{2}d_9         \\
			           &            & -(d_2+d_3) & \sqrt{2}d_5        & \sqrt{2}d_7        & -\sqrt{2}(d_8 +d_9) \\\hline
			           &            &            &                    &                    &                     \\
			\completesymmatrix{}                                                                                 \\
			           &            &            &                    &                    &                     \\
		\end{array}
		\right]
	\mathbf{B}^{\mathbf{v}}_{\xi} \otimes \mathbf{B}^{\mathbf{v}}_{\zeta}
	}
	\nonumber,
\end{align}
the operator $\text{sym}(\cdot)$ projecting onto the completely symmetric part of a tensor
and the operator $\text{dev}(\cdot)$ extracting the deviatoric part of a tensor~\cite{Spencer1970}.
For clarity, the constant and second-order part of equation \eqref{eq:N4_simplified_triclinic}
is given explicitly
\begin{align}
	&
	\A^{\text{iso}}
	+
	\frac{6}{7} \;
	\text{sym}
	\left(
	\text{dev}\left(\mathbf{A}(\lambda_1\,, \lambda_2)\right)
	\otimes
	\Id
	\right)
	=
	\label{eq:constant_and_second_order_part_of_triclinic_parameterization}\\
	&\quad
	{
	\left[
		\begin{array}{lll|lll}
			\frac{6}{7} \lambda_1 - \frac{3}{35}  & \frac{1}{7} \lambda_1 + \frac{1}{7} \lambda_2 - \frac{1}{35}        & -\frac{1}{7} \lambda_2 + \frac{4}{35}        & 0 & 0 & 0 \\
			           & \frac{6}{7} \lambda_2 - \frac{3}{35} & -\frac{1}{7} \lambda_1 + \frac{4}{35}       & 0 & 0 & 0 \\
			           &            & - \frac{6}{7} \lambda_1 - \frac{6}{7} \lambda_2 + \frac{27}{35} & 0 & 0 & 0 \\\hline
			           &            &            &                    &                    &                     \\
			\completesymmatrix{}                                                                                 \\
			           &            &            &                    &                    &                     \\
		\end{array}
		\right]
	\mathbf{B}^{\mathbf{v}}_{\xi} \otimes \mathbf{B}^{\mathbf{v}}_{\zeta}
	}
	\nonumber,
\end{align}
already including the normalization constraint $\lambda_3 = 1- \lambda_1 - \lambda_2$
enforced by the condition \eqref{eq:fot_zeroth_one}.
If we follow the ordering convention on the eigenvalues of $\mathbf{A}$ in expression \eqref{eq:convention_eigenvalues_N2}
and decide for either right- or left-handed coordinates,
a symmetric second-order tensor, which is automatically orthotropic, {will have} at least four eigensystems.
These four systems differ by the action of elements of the orthotropic symmetry group, each changing signs of two eigenvectors at a time.
Without loss of generality, we choose one of the possible four eigensystems to be defined by
\begin{equation}
	\mathbf{Q} = \sum_{i=1}^3 \mathbf{v}_i \otimes \mathbf{e}_i,
	\label{eq:mapping}
\end{equation}
mapping
the arbitrary but fixed basis $\left\{\mathbf{e}_i\right\}$
onto the eigensystem $\left\{\mathbf{v}_i\right\}$.
Selecting one of the other three possible eigensystems results in two sign changes of groups of tensor coefficients
but leaves the identified physical quantity unaffected.
The Kelvin-Mandel~\cite{Thomson1856, Mandel1965} basis $\mathbf{B}^{\mathbf{v}}_{\xi} \otimes \mathbf{B}^{\mathbf{v}}_{\zeta}$
in \eqref{eq:triclinic_structure_tensor}
and \eqref{eq:constant_and_second_order_part_of_triclinic_parameterization}
is spanned in the eigensystem and therefore, e.g.,
$\boldsymbol{B}_1  = \mathbf{v}_1 \otimes \mathbf{v}_1$ holds.
This simplifies incorporating specific material symmetry.

\subsection{Material symmetry constraints}\label{sec:material_symmetry_constraints}
Constraints of material symmetry reduce the dimensionality of the optimization problem stated in
equation \eqref{eq:semidefinite_program_tensor}.
Furthermore, material symmetry constraints arise naturally in the context of closure approximations,
as any second-order fiber-orientation tensor is (at least) orthotropic,
whereas fourth-order fiber-orientation tensors may not be orthotropic, in general.
A rotation
$\mathbf{Q} \in \specialorthogonalgroup{d}$
is said to belong to the symmetry class $\symmetryclass{\N}$
of a {given tensor} $\N$,
{provided its action by rotation leaves the tensor invariant}, i.e.,
the characterization
\begin{equation}
	\symmetryclass{\N}
	=
	\left\{
		\mathbf{Q} \in \specialorthogonalgroup{d} \,\middle|\, \mathbf{Q} \star \N	= \N
	\right\}
\end{equation}
holds with the Rayleigh product $\star$ rotating each tensor basis of the tensor $\N$ by the rotation $\mathbf{Q}$.
Forte and Vianello~\cite{Forte1996} identified the possible symmetry classes of fourth-order tensors,
which read - ordered by increasing cardinality of the symmetry group $\symmetryclass{\N}$ -
triclinic, monoclinic, orthotropic, trigonal, tetragonal, transversely isotropic, cubic and isotropic.
Isotropic quantities are symmetric with respect to all rotations,
whereas the triclinic symmetry group is {trivial,} $\symmetryclass{\N} = \{\Id\}$.

Material symmetry constraints impose linear constraints in the optimization problem \eqref{eq:semidefinite_program_tensor}.
For example, orthotropic symmetry w.r.t. the symmetry axes in three-dimensions is equivalent to vanishing parameters
$d_i=0$ for $i \in \{4,\,5,\,6,\,7,\,8,\,9\}$ in the parameterization
\eqref{eq:N4_simplified_triclinic},
as the orthotropic symmetry class
\begin{equation}
	\symmetryclassspecial{\text{orthotropic}}
	=
	\left\{
	\mathbf{S}^{\text{ortho}}_1,
	\mathbf{S}^{\text{ortho}}_2,
	\mathbf{S}^{\text{ortho}}_3,
	\mathbf{S}^{\text{ortho}}_4
	\right\}
\end{equation}
consists of the rotations
\begin{align}
	& \mathbf{S}^{\text{ortho}}_1 = \left[
	   \begin{array}{rrr}
		   \phantom{-}1 & 0 & 0 \\
		   0 & \phantom{-}1 & 0 \\
		   0 & 0 & \phantom{-}1
	   \end{array}
	   \right]
   \;\mathbf{v}_i \otimes \mathbf{v}_j,
	& \mathbf{S}^{\text{ortho}}_2 = \left[
	   \begin{array}{rrr}
		   \phantom{-}1 & 0  & 0  \\
		   0 & -1 & 0  \\
		   0 & 0  & -1
	   \end{array}
	   \right]
   \;\mathbf{v}_i \otimes \mathbf{v}_j,
   \\
	& \mathbf{S}^{\text{ortho}}_3 = \left[
	   \begin{array}{rrr}
		   -1 & 0 & 0  \\
		   0  & \phantom{-}1 & 0  \\
		   0  & 0 & -1
	   \end{array}
	   \right]
   \;\mathbf{v}_i \otimes \mathbf{v}_j,
	& \mathbf{S}^{\text{ortho}}_4 = \left[
	   \begin{array}{rrr}
		   -1 & 0  & 0 \\
		   0  & -1 & 0 \\
		   0  & 0  & \phantom{-}1
	   \end{array}
	   \right]
   \;\mathbf{v}_i \otimes \mathbf{v}_j,
\end{align}
expressed in a coordinate system $\bigl\{ \mathbf{v}_i \bigr\}$
whose axes are normal to symmetry planes
of the orthotropic quantity.

\subsection{Extreme fourth-order information}\label{sec:extreme_fourth_order_information}
We are interested in the phase space of fourth-order fiber-orientation tensors,
extending previous work of Bauer and Böhlke~\cite{Bauer2022Variety}.
For this purpose, we fix a second-order fiber-orientation tensor and
study the set of all fourth-order fiber-orientation tensors
which cover the prescribed second-order tensor
in the sense that the condition $\A : \Id = \mathbf{A}$ holds.
This set is closed and convex, and we wish to study the ``extreme states'' of this set.
For this purpose and with the notation introduced in the previous sections,
suppose that a fiber-orientation tensor $\mathbf{A}(\lambda_1,\, \lambda_2)$
is given in its principal axes and where $\lambda_1,\, \lambda_2$ refer to the two largest eigenvalues.
Then, we consider an arbitrary direction
\begin{equation}
	\mathbf{n}(\varphi, \theta)
	=
	\left[
		\begin{array}{l}
			\cos(\varphi) \sin(\theta)\\
			\sin(\varphi) \sin(\theta)\\
			\cos(\theta)
		\end{array}
		\right]
	\;\mathbf{v}_i
\end{equation}
on the unit sphere, parameterized by the angles $\varphi$ and $\theta$.
We consider {the tensor} $\C = \mathbf{n}^{\otimes4}$ in problem \eqref{eq:semidefinite_program_tensor}
combined with the constraint $\A : \Id = \mathbf{A}(\lambda_1,\, \lambda_2)$, specifying the second-order information.
In addition, material symmetry is ensured by enforcing $\mathbf{Q} \star \A$
for all rotations $\mathbf{Q} \in \symmetryclassspecial{\text{problem}}$
of the selected symmetry class $\symmetryclassspecial{\text{problem}}$,
leading to the problem
\begin{equation}
	\begin{array}{ll}
		\text{maximize}  		& \mathbf{n}^{\otimes4} :: \A\\
		\text{subject to}		& \A \in \candidateSet{3},\\
								& \A : \Id = \mathbf{A}(\lambda_1,\, \lambda_2),\\
								& \mathbf{Q} \star \A = \A \quad \forall \, \mathbf{Q} \in \symmetryclassspecial{\text{problem}}.
	\end{array}
	\label{eq:program_max_fourth_order_info}
\end{equation}
\newcommand{\mywidth}{105mm}


\begin{figure}[!t]
	\centering
	\includegraphics[width=\mywidth/2]{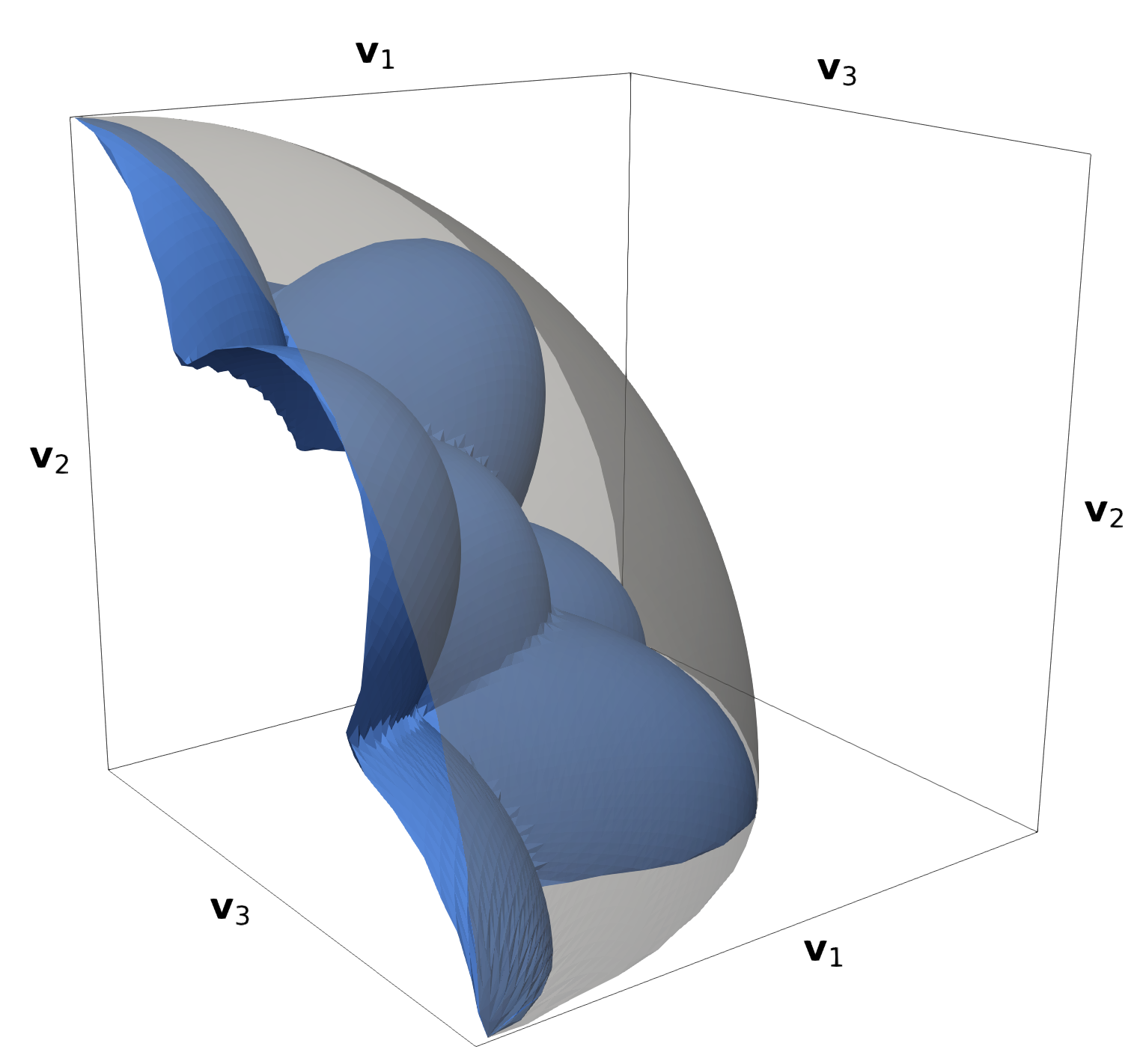}
	\includegraphics[width=\mywidth/2]{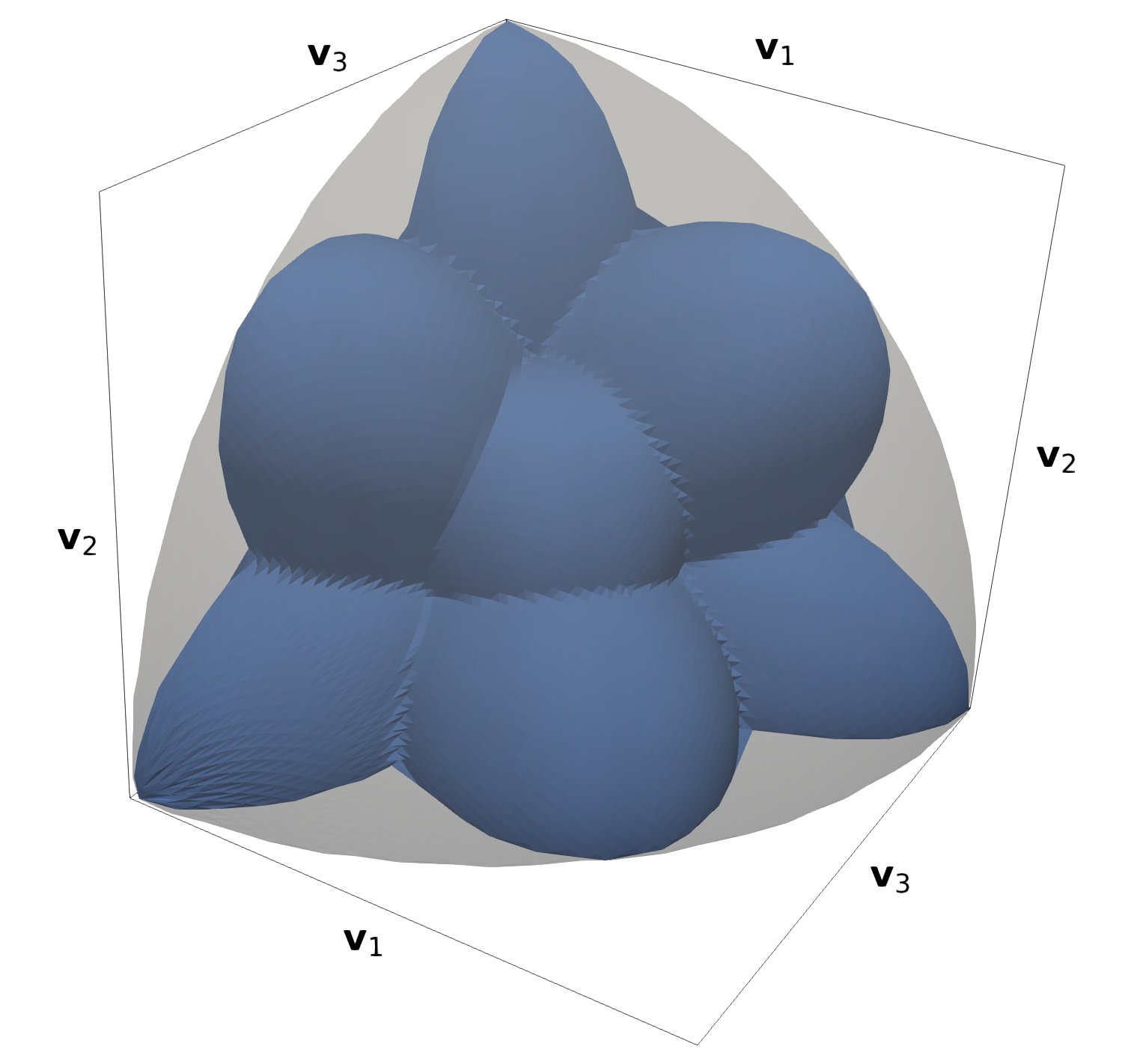}
	\caption{%
		Two views on identical surfaces of
		polar plots of the maximum objective function value,
		i.e., $\mathbf{n}^{\otimes4}(\varphi, \theta=0) :: \A$,
		of problem \eqref{eq:program_max_fourth_order_info}
		for values $\lambda_1 = 1/3$ and $\lambda_2 = 1/3$.
		Results for the unconstrained triclinic case, i.e.,
		$\symmetryclassspecial{\text{problem}} = \symmetryclassspecial{\text{triclinic}}$
		are given in light gray, whereas
		results with restriction to the orthotropic subspace, i.e.,
		$\symmetryclassspecial{\text{problem}} = \symmetryclassspecial{\text{orthotropic}}$
		are shown in blue.
	}
	\label{fig:sphere_1}
\end{figure}

\begin{figure}[!t]
	\centering
	\includegraphics[width=\mywidth]{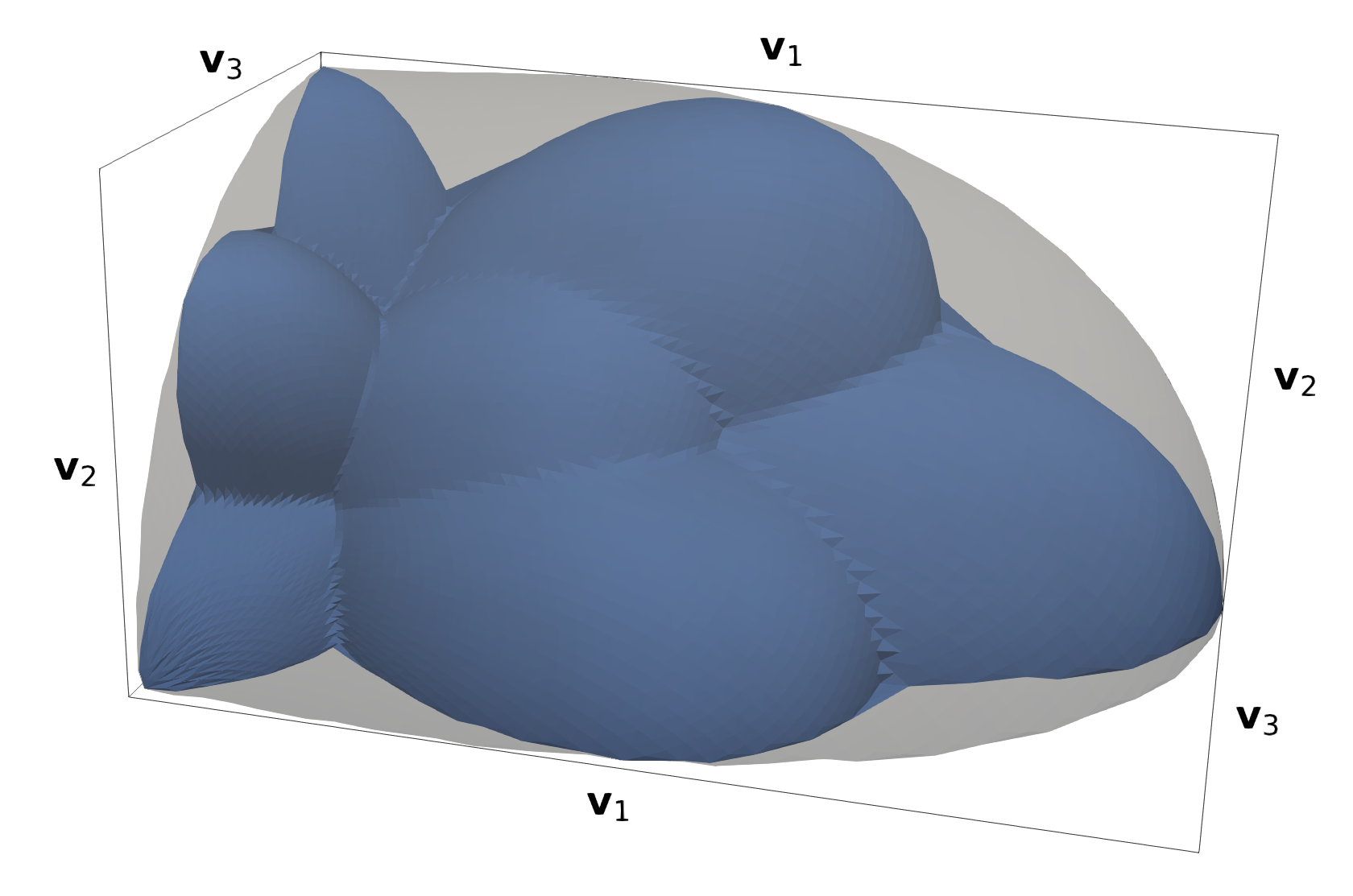}
	\caption{%
		Polar plots of the maximum objective function value,
		i.e., $\mathbf{n}^{\otimes4}(\varphi, \theta=0) :: \A$,
		of problem \eqref{eq:program_max_fourth_order_info}
		for values $\lambda_1 = 1/2$ and $\lambda_2 = 1/4$.
		Results for the unconstrained triclinic case, i.e.,
		$\symmetryclassspecial{\text{problem}} = \symmetryclassspecial{\text{triclinic}}$
		are given in light gray, whereas
		results with restriction to the orthotropic subspace, i.e.,
		$\symmetryclassspecial{\text{problem}} = \symmetryclassspecial{\text{orthotropic}}$
		are shown in blue.
	}
	\label{fig:sphere_2}
\end{figure}

\begin{figure}[!b]
	\centering
	\includegraphics[width=\mywidth]{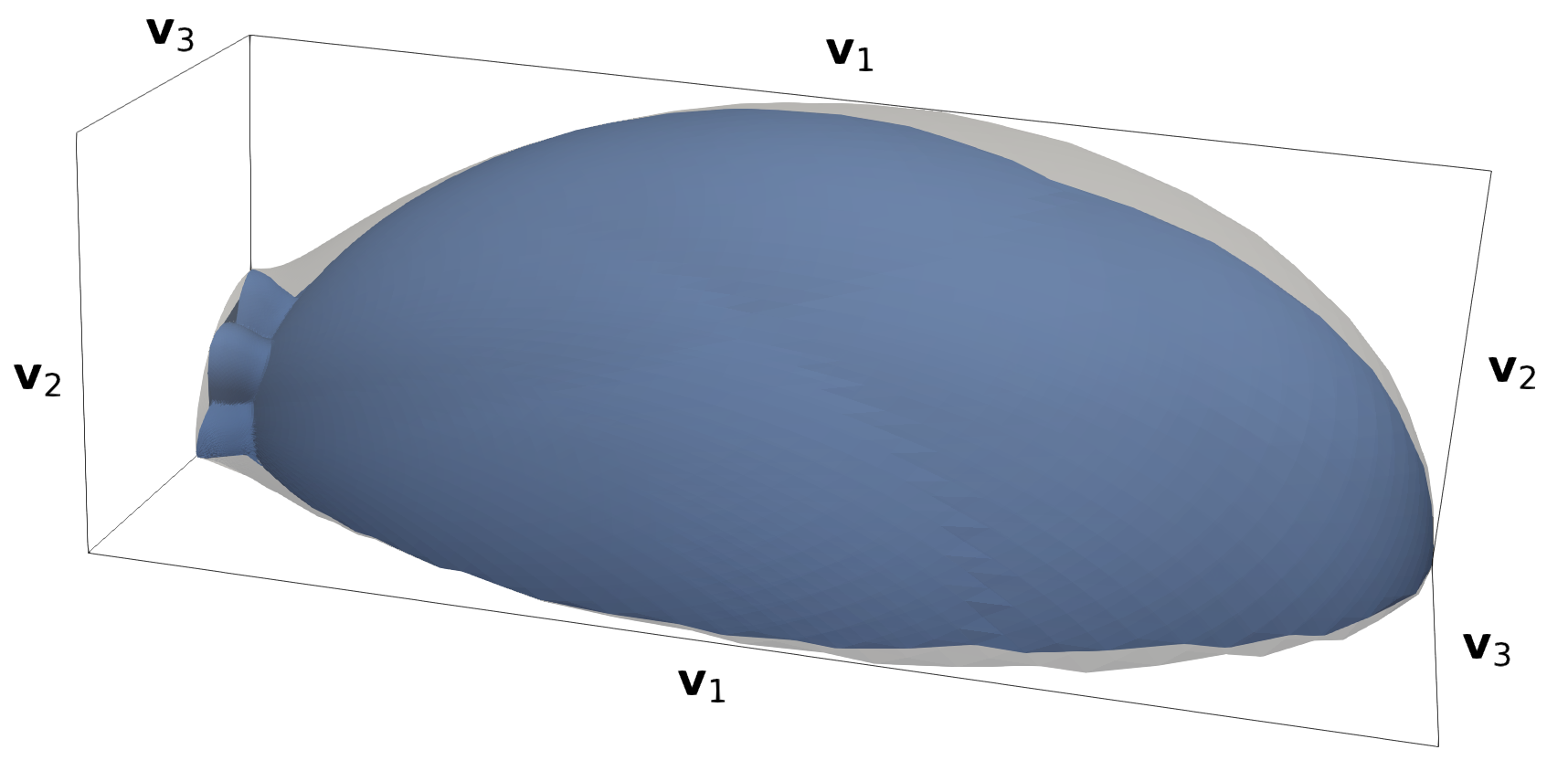}
	\caption{%
		Polar plots of the maximum objective function value,
		i.e., $\mathbf{n}^{\otimes4}(\varphi, \theta=0) :: \A$,
		of problem \eqref{eq:program_max_fourth_order_info}
		for values $\lambda_1 = 8/10$ and $\lambda_2 = 1/10$.
		Results for the unconstrained triclinic case, i.e.,
		$\symmetryclassspecial{\text{problem}} = \symmetryclassspecial{\text{triclinic}}$
		are given in light gray, whereas
		results with restriction to the orthotropic subspace, i.e.,
		$\symmetryclassspecial{\text{problem}} = \symmetryclassspecial{\text{orthotropic}}$
		are shown in blue.
	}
	\label{fig:sphere_3}
\end{figure}
%
%
%
For the convenience of the reader,
we translated typical constraints
of problem \eqref{eq:program_max_fourth_order_info}
in terms of {expressions} $\GG_k :: \A = g_k$ with $k = 1, ..., m$
with {tensors} $\GG_k = G^k_{\xi \zeta} \, \mathbf{B}^{\mathbf{v}}_{\xi} \otimes \mathbf{B}^{\mathbf{v}}_{\zeta}$,
see \autoref{tab:constraints} in appendix \ref{app:constraints}.
%

\subsection{Results in three spatial dimensions}\label{sec:results_3d}
On a regular grid of points on one-eighth of the unit sphere, the maximum values of the objective function
of problem \eqref{eq:program_max_fourth_order_info}
are visualized in Figures \ref{fig:sphere_1} to \ref{fig:sphere_3}
as spherical plots.
Due to the symmetry of the objective function specified by the quartic product $\mathbf{n}^{\otimes4}$,
inspecting the results on a grid of direction within one-eighth of the unit sphere is sufficient.
Figures \ref{fig:sphere_1} to \ref{fig:sphere_3} differ by the given second-order fiber-orientation information, i.e., tuples
$\mathbf{A}(\lambda_1, \lambda_2)$.
Each of the figures contains two surfaces.
The first surface plotted in light gray represents the maximum objective function obtained without material constraints on $\A$,
whereas the second surface plotted in blue represents
the same quantity
with $\A$ restricted to
orthotropic fourth-order fiber-orientation tensors, i.e.,
$\symmetryclassspecial{\text{problem}} = \symmetryclassspecial{\text{orthotropic}}$
in problem \eqref{eq:program_max_fourth_order_info}.
Figures \ref{fig:sphere_1} to \ref{fig:sphere_3} demonstrate that restricting to the orthotropic subspace
limits the possibility to obtain large values of the objective function in specific directions.
Only in directions aligned with the axes of the eigensystem $\bigl\{ \mathbf{v}_i \bigr\}$
as well as
in directions with
$(\varphi,\, \theta) \in \bigl[(0,\, 45^\circ),\, (90^\circ,\, 45^\circ),\, (45^\circ,\, 0)\bigr]$,
the constraint of orthotropy does not induce a restriction on the extreme value of the objective function.
With increasing anisotropy of the specified second-order information,
the
consequences of the constraint onto the orthotropic subspace decrease.
%
%

\subsection{Results in two spatial dimensions}\label{sec:results_2d}
If the given second-order fiber-orientation $\mathbf{A}$ is planar,
i.e., fulfills the conditions $\lambda_2 = 1-\lambda_1$ and $1/2 \le \lambda_1 \le 1$,
the constraint $\A : \Id = \mathbf{A}(\lambda_1,\, \lambda_2)$ in problem \eqref{eq:program_max_fourth_order_info}
{will enforce} planarity of the resulting fourth-order fiber-orientation tensor $\A$.
Within the plane spanned by the directions $\mathbf{v}_1$ and $\mathbf{v}_2$,
the maximum values of the objective function $\mathbf{n}^{\otimes4} :: \A$
are plotted in
Figures \ref{fig:planar_01} to \ref{fig:planar_06} with and without
constraints in terms of material symmetry.
The directions $\mathbf{n}(\varphi, \theta)$
within the plane $\mathbf{v}_1$ and $\mathbf{v}_2$ are characterized by the condition $\theta=0$
and varying angle $\varphi$.
The values of the objective function obtainable for directions aligned with $\mathbf{v}_1$ and $\mathbf{v}_2$ are
$\lambda_1$ and $\lambda_2$, respectively.
For directions specified by {angles}
$\theta \in \{0,\, 45^\circ,\, 90^\circ\}$,
the objective values obtained
with or without material symmetry constraints are identical.
For all other directions,
restricting to the class of orthotropic fiber-orientation tensors $\A$
limits the values of the objective function that can be attained.
This limitation decreases as second-order fiber-orientation tensor information advances towards the uni-directional state, i.e., $\lambda_1 \rightarrow 1$.

\renewcommand{\mywidth}{50mm}
\newcommand{\mytikzwidth}{47mm}
\begin{figure}[h]
	\begin{minipage}{\mywidth}
		\centering
		\includegraphics[scale=1]{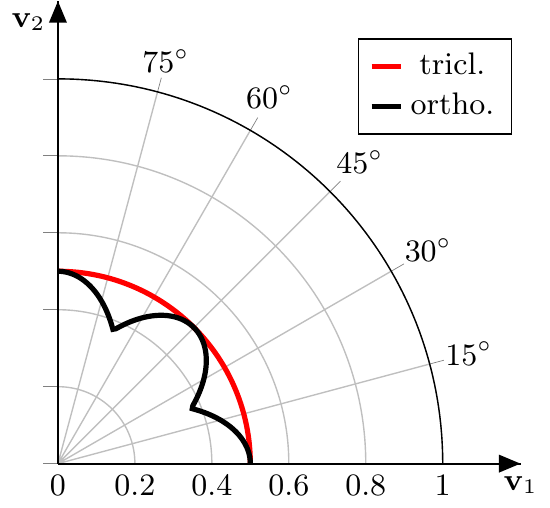}
		\caption{$\lambda_1 = 5/10$}
		\label{fig:planar_01}
	\end{minipage}%
	\hfill
	\begin{minipage}{\mywidth}
		\centering
		\includegraphics[scale=1]{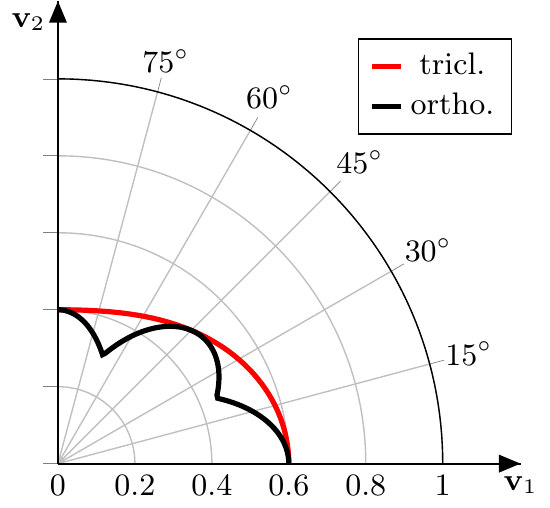}
		\caption{$\lambda_1 = 6/10$}
		\label{fig:planar_02}
	\end{minipage}%
	\hfill
	\begin{minipage}{\mywidth}
		\centering
		\includegraphics[scale=1]{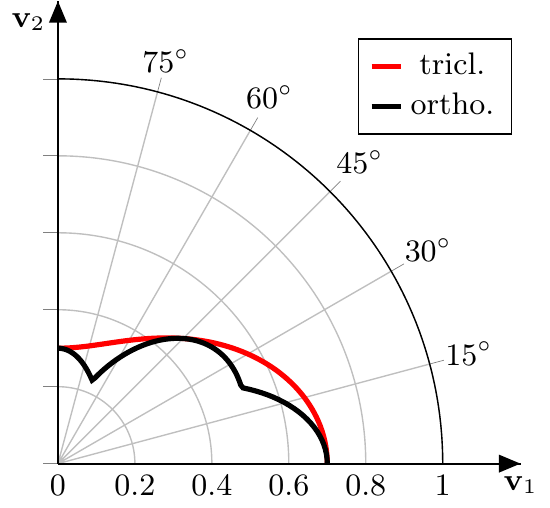}
		\caption{$\lambda_1 = 7/10$}
		\label{fig:planar_03}
	\end{minipage}

	\begin{minipage}{\mywidth}
		\centering
		\includegraphics[scale=1]{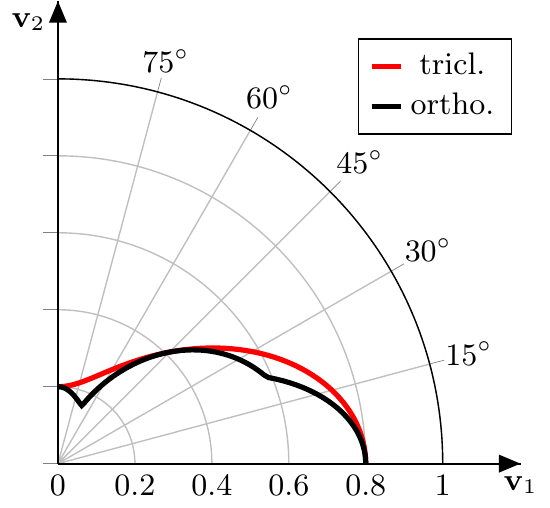}
		\caption{$\lambda_1 = 8/10$}
		\label{fig:planar_04}
	\end{minipage}%
	\hfill
	\begin{minipage}{\mywidth}
		\centering
		\includegraphics[scale=1]{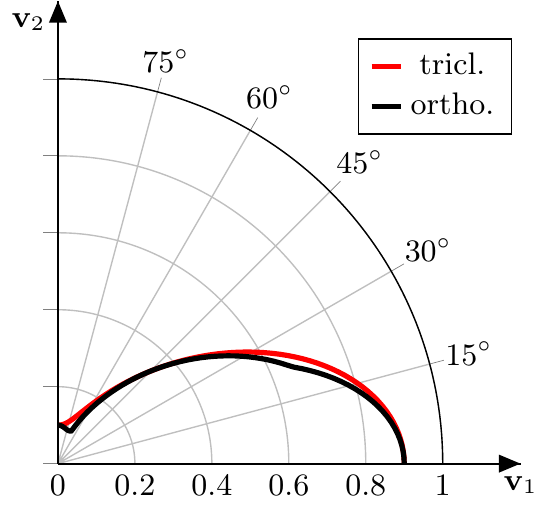}
		\caption{$\lambda_1 = 9/10$}
		\label{fig:planar_05}
	\end{minipage}%
	\hfill
	\begin{minipage}{\mywidth}
		\centering
		\includegraphics[scale=1]{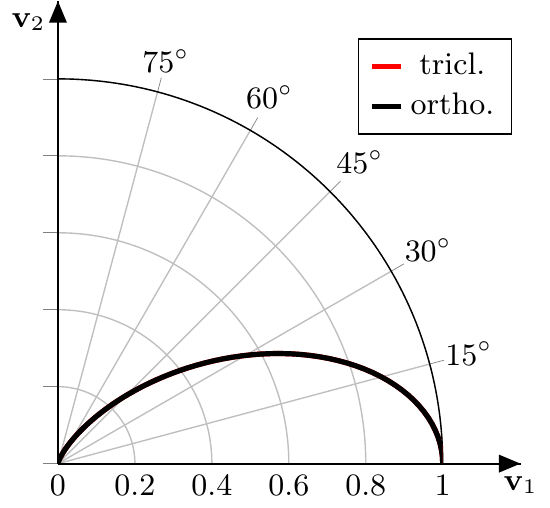}
		\caption{$\lambda_1 = 1$}
		\label{fig:planar_06}
	\end{minipage}%
	\caption{Polar plots of the maximum objective function value,
	i.e., $\mathbf{n}^{\otimes4}(\varphi, \theta=0) :: \A$,
	of problem \eqref{eq:program_max_fourth_order_info}
	and $\lambda_2 = 1- \lambda_1$ for selected values of $\lambda_1$.
	Results are given for restriction to the triclinic or orthotropic subspace, i.e.,
	$\symmetryclassspecial{\text{problem}} = \symmetryclassspecial{\text{triclinic}}$
	or
	$\symmetryclassspecial{\text{problem}} = \symmetryclassspecial{\text{orthotropic}}$
	respectively.
	}
\end{figure}

\section{Conclusion}
\label{sec:conclusion}

{The work at hand showed that} the set of candidates for fourth-order fiber-orientation tensors
\begin{equation}
	\candidateSet{d} = \left\{ \A \in \Sym{d}{4} \, \middle| \,
	S:\A:S \geq 0 \quad \text{for all} \quad S \in \Sym{d}{2}, \quad \Id:\A:\Id =
	1\right\}
	\nonumber
\end{equation}
{represents all} realizable {fourth-order} fiber-orientation tensors for {the practically relevant spatial dimensions} $d=2$ or $d=3$. {This result provides an a-posteriori justification for} the approach of Bauer and Böhlke~\cite{Bauer2022Variety} {to characterize the variety of fourth-order fiber-orientation tensors based on tensor decompositions.\\
Our result has a number of interesting implications and applications. For a start, it permits assessing the realizability of a given fourth-order tensor as a fiber-orientation tensor in a simple and straightforward way. Consider, for instance, fiber-microstructure generators ~\cite{Schneider2017,Mehta2022,CurvedFibers2022}, which serve as the basis for full-field simulations and where the fourth-order fiber-orientation tensor serves as the input. Then, if the prescribed tensor does not belong to the set $\candidateSet{d}$, it \emph{cannot be realized} by a fiber microstructure at all, independent of the size of the microstructure and the number of considered fibers. This simple fact provides constraints on fiber-orientation closure approximations used in fiber-orientation dynamics~\cite{Cintra1995,Chung2001}, and may be used to re-evaluate their physical plausibility.\\
A second consequence concerns the effective elastic properties of short-fiber reinforced composites. It is well-known that, together with the volume fraction and the aspect ratio, the fourth-order fiber-orientation tensor is responsible for the effective elastic properties of such a composite~\cite{Jack2007,Jack2008,Mueller2016}. Classical closure approximations, expressing the fourth-order fiber-orientation tensor as {an isotropic tensor} function of the second-order fiber-orientation tensor, are restricted, {by definition}, to orthotropic symmetry. In particular, the set of realizable effective stiffnesses is restricted to being orthotropic. Our computational investigations revealed that this restriction to orthotropy actually underestimates the achievable effective stiffnesses in particular directions for a prescribed second-order fiber-orientation tensor significantly. Thus, we expect that using the full phase space of fourth-order fiber-orientation tensors will be critical for exploiting the full lightweight potential of such short-fiber composites, e.g., by reducing safety factors. First steps in this direction were already proposed~\cite{AdvaniTucker1987,Altan1990,Jack2006_6thOrderClosure}, and these works may be revisited with fresh impetus.\\
Last but not least let us stress the simplicity and utility of the connection between optimization of fiber microstructures described by fourth-order fiber-orientation tensors and semi-definite programming introduced in the work at hand. In particular due to the power of the available optimization solvers, new directions were opened to engineers working on fiber-reinforced components and their optimal design.
}

\section*{Acknowledgements}

The authors would like to thank J. Köbler (ITWM) and M. Krause (KIT) for fruitful discussions.
Support by the German Research Foundation (DFG, Deutsche Forschungsgemeinschaft)
within the International Research Training Group
``Integrated engineering of continuous-discontinuous long fiber reinforced polymer structures''
(GRK 2078/2) - project 255730231 - is gratefully acknowledged.
Partial support of MS by the European Research Council within the
Horizon Europe program - project 101040238 - is gratefully acknowledged.

\section*{Declarations}

\subsection*{Funding}
JKB received funding by German Research Foundation (DFG, Deutsche Forschungsgemeinschaft)
within the International Research Training Group
``Integrated engineering of continuous-discontinuous long fiber reinforced polymer structures''
(GRK 2078/2) - project 255730231.
MS received partial support by the European Research Council within the
Horizon Europe program - project 101040238.

\subsection*{Competing interests}
The authors have no competing interests to declare that are relevant to the content of this article.

\appendix

\section{Representing a fiber-orientation tensor as a non-negative sum of
  monomials}
\label{sec:atomic_decomposition_Ak}

The purpose of this appendix is to show the monomial decomposition
\eqref{sec:fot_def_monomial_decomposition}, i.e., that every fixed
fiber-orientation tensor $A^{(k)}$ of even order may be written as a finite sum
\begin{equation}\label{eq:atomic_decomposition_Ak_monomial_decomposition}
	A^{(k)} = \sum_{i=1}^r w_i \, q_i^{\otimes k}
\end{equation}
with appropriate non-negative weights $w_i$, directions $q_i \in S^{d-1}$ and a
non-negative integer $r$ not exceeding ${{k+d-1}\choose{d-1}}$, the dimension
of the space of completely symmetric $k$-tensors in $d$ dimensions.\\
As a prerequisite, we require Carathéodory's theorem~\cite{Caratheodory1911},
which states that any vector $z$ in an $m$-dimensional vector space $V$ which
is represented in the form
\begin{equation}\label{eq:atomic_decomposition_Ak_Caratheodory_original}
	z = \sum_{i=1}^r x_i, \quad x_i \in V,
\end{equation}
may also be written in the form
\begin{equation}\label{eq:atomic_decomposition_Ak_Caratheodory}
	z = \sum_{i \in \mathcal{I}} \rho_i\,x_i
\end{equation}
for an index set $\mathcal{I} \subseteq \{1,\ldots,r\}$ with at most $m$
elements and non-negative coefficients $\rho_i$. For the convenience of the
reader, we include a short derivation, following Reznick~\cite{Reznick}. Let us
consider the case
\begin{equation}\label{eq:atomic_decomposition_Ak_Caratheodory_special}
	z = \sum_{i=1}^{m+1} x_i, \quad \text{i.e.}, \quad r=m+1.
\end{equation}
The vectors $\{x_i\}$ are linearly dependent, i.e., there are coefficients
$c_i$, not all zero, s.t. the representation
\begin{equation}\label{eq:atomic_decomposition_Ak_linear_dependence}
	0 = \sum_{i=1}^{m+1} c_i \, x_i
\end{equation}
holds. We may assume that some of the $c_i$ are positive. Indeed, if they were
non-positive, we may multiply the coefficients by $-1$. Furthermore, suppose
that the coefficients $c_i$ are ordered, upon possibly reindexing the sequence.
In particular, the inequality
\begin{equation}\label{eq:atomic_decomposition_Ak_nonnegative_coefficients}
	c_i \leq c_{m+1}, \quad \text{i.e.}, \quad 0 \leq 1 -
	\frac{c_i}{c_{m+1}},
\end{equation}
holds for all $i=1,\ldots,m$ as $c_{m+1}>0$. We may rearrange equation
\eqref{eq:atomic_decomposition_Ak_linear_dependence} into the form
\begin{equation}
	x_{m+1} = - \sum_{i=1}^m \frac{c_i}{c_{m+1}} \, x_i.
\end{equation}
Inserting this representation into the expression
\eqref{eq:atomic_decomposition_Ak_Caratheodory_special} yields
\begin{equation}
	z = \sum_{i=1}^m x_i + x_{m+1} = \sum_{i=1}^m \left( 1 -
	\frac{c_i}{c_{m+1}} \right) \, x_i.
\end{equation}
Thus, we arrived at the conclusion of Carathéodory's theorem for the special
case \eqref{eq:atomic_decomposition_Ak_Caratheodory_special} in view of the
non-negativity assertion
\eqref{eq:atomic_decomposition_Ak_nonnegative_coefficients} of the
coefficients.\\
The general case \eqref{eq:atomic_decomposition_Ak_Caratheodory_original} with
more terms $r>m+1$ follows from applying the special case
\eqref{eq:atomic_decomposition_Ak_Caratheodory_special} inductively.\\
Suppose a fiber-orientation tensor $A^{(k)}$ is given, i.e., there is some
probability measure $\mu$ and a representation \eqref{eq:fot_def}
\begin{equation}\label{sec:atomic_decomposition_Ak_fot_def}
	A^{(k)} = \int_{S^{d-1}} p^{\otimes k} \, d\mu(p).
\end{equation}
The space of Radon measures may be considered as the continuous dual space of
the space of continuous functions on the unit sphere $S^{d-1}$. It is a
classical result of functional analysis
(e.g., as a direct consequence of the Krein-Milman theorem~\cite[Example 8.16]{Simon2011}
)
that the sum of Dirac
measures is dense in the space of Radon measures w.r.t. the weak-$*$ topology,
i.e., there are
$r_{j}$ positive weights $w_{i,j}$ and directions
$p_{i,j} \in S^{d-1}$, s.t., for any continuous function $\phi:S^{d-1}
	\rightarrow \R$, we have
\begin{equation}
	\sum_{i=1}^{r_{j}} w_{i,j} \, \phi(p_{i,j}) \rightarrow \int_{S^{d-1}}
	\phi(p)\, d\mu(p) \quad \text{as} \quad j \rightarrow \infty.
\end{equation}
Choosing the monomials of homogeneity four as special continuous functions, we
thus obtain the result
\begin{equation}
	\sum_{i=1}^{r_{j}} w_{i,j} \, p_{i,j}^{\otimes k} \rightarrow A^{(k)}
	\quad \text{as} \quad j \rightarrow \infty.
\end{equation}
By Carathéodory's theorem \eqref{eq:atomic_decomposition_Ak_Caratheodory} we
can assume a uniform bound on the ranks $r_j$, i.e., it holds
\begin{equation}
	\sum_{i=1}^{r} w_{i,j} \, p_{i,j}^{\otimes k} \rightarrow A^{(k)} \quad
	\text{as} \quad j \rightarrow \infty.
\end{equation}

\section{Constraints}\label{app:constraints}
Typical constraints
of problem \eqref{eq:program_max_fourth_order_info}
are translated
in terms of the condition $\GG_k :: \A = g_k$ with $k = 1, ..., m$
with $\GG_k = G^k_{\xi \zeta} \, \mathbf{B}^{\mathbf{v}}_{\xi} \otimes \mathbf{B}^{\mathbf{v}}_{\zeta}$
in \autoref{tab:constraints}.
\begin{table}[!ht]
	\centering
	\begin{tabular}{l|l|l}
		Description &
		Non-vanishing coefficients $G^k_{\xi \zeta}$ &
		$g_k$ \\\hline\hline
		Complete symmetry of $\A$ & $G_{44} = 1,\, G_{23} = -2$ & 0	\\
		                          & $G_{55} = 1,\, G_{13} = -2$ & 0	\\
		                          & $G_{66} = 1,\, G_{12} = -2$ & 0	\\
		                          & $G_{45} = 1,\, G_{36} = -\sqrt{2}$ & 0	\\
		                          & $G_{46} = 1,\, G_{25} = -\sqrt{2}$ & 0	\\
		                          & $G_{56} = 1,\, G_{14} = -\sqrt{2}$ & 0	\\\hline
		$\A$ in eigensystem of $\mathbf{A}$ & $G_{14} = G_{24} = G_{34} = 1$ & 0\\
		                                    & $G_{15} = G_{25} = G_{35} = 1$ & 0\\
		                                    & $G_{16} = G_{26} = G_{36} = 1$ & 0 \\\hline
		Normalization			& $G_{11} = G_{22} = G_{33} = G_{44} = G_{55} = G_{66} = 1$ & 1 \\\hline
		Eigenvalues of $\A : \Id$   & $G_{11} = G_{12} = G_{13} = 1$ & $\lambda_1$ \\
									& $G_{12} = G_{22} = G_{23} = 1$ & $\lambda_2$ \\\hline
		$\symmetryclassspecial{\text{problem}} = \symmetryclassspecial{\text{orthotropic}}$
						& $G_{24} = 1$ & 0\\
						& $G_{34} = 1$ & 0\\
						& $G_{15} = 1$ & 0\\
						& $G_{35} = 1$ & 0\\
						& $G_{16} = 1$ & 0\\
						& $G_{26} = 1$ & 0
	\end{tabular}
	\caption{Constraints of problem \eqref{eq:program_max_fourth_order_info}
	reformulated in terms of the condition $\GG_k :: \A = g_k$
	with $\GG_k = G^k_{\xi \zeta} \, \mathbf{B}^{\mathbf{v}}_{\xi} \otimes \mathbf{B}^{\mathbf{v}}_{\zeta}$}
	\label{tab:constraints}
\end{table}

\bibliographystyle{ieeetr}
\bibliography{literature}

\end{document}